\documentclass[conference]{IEEEtran}
\usepackage{amsfonts}
\usepackage{graphicx}
\usepackage{algorithm}
\usepackage{algorithmic}
\usepackage{amssymb}
\usepackage{amsmath}
\usepackage{setspace}
\usepackage{xcolor}
\usepackage{graphicx}
\usepackage{subfigure}
\usepackage{lipsum}

\begin{document}
\title{Dynamic Channel Access via Meta-Reinforcement Learning}

\author{\IEEEauthorblockN{Ziyang Lu and M. Cenk Gursoy}
	\IEEEauthorblockA{Department of Electrical Engineering and Computer Science, Syracuse University, Syracuse, NY
		\\
		Email:zlu112@syr.edu, mcgursoy@syr.edu}}

\maketitle

\begin{abstract}
In this paper, we address the channel access problem in a dynamic wireless environment via meta-reinforcement learning. Spectrum is a scarce resource in wireless communications, especially with the dramatic increase in the number of devices in networks. Recently, inspired by the success of deep reinforcement learning (DRL), extensive studies have been conducted in addressing wireless resource allocation problems via DRL. However, training DRL algorithms usually requires a massive amount of data collected from the environment for each specific task and the well-trained model may fail if there is a small variation in the environment. In this work, in order to address these challenges, we propose a meta-DRL framework that incorporates the method of Model-Agnostic Meta-Learning (MAML). In the proposed framework, we train a common initialization for similar channel selection tasks. From the initialization, we show that only a few gradient descents are required for adapting to different tasks drawn from the same distribution. We demonstrate the performance improvements via simulation results.
\end{abstract}

\IEEEpeerreviewmaketitle

\section{Introduction}
\subsection{Background}
\subsubsection{Channel Selection}
Spectrum has become a precious resource in wireless networks due to the increase in device density and network scale. Hence, how to efficiently allocate the limited number of channels is critical and has been extensively studied recently. Additionally, inspired by the success of deep reinforcement learning (DRL) in decision making in dynamic environments (e.g., in gaming \cite{mnih2013playing}) and its model-free nature, DRL has already been applied to address wireless channel access problems for instance, recently in \cite{wang2018deep}, \cite{zhong2019deep}, \cite{naparstek2018deep}, where a model is trained for solving a specific channel selection task. DRL demonstrates near-optimal performance in these studies. However, training such a model requires large quantities of data collected from the environment, which can be expensive in wireless networks. Besides, the models are usually task-specific and they can potentially fail when the task or the environment tends to vary.
\subsubsection{Meta-Reinforcement Learning}
Humans can learn quickly in an unseen task if they have experience in similar tasks before. Inspired by this fact, the concept of meta-learning has been proposed, where the algorithm can automatically find hyperparameters and architectures of the models for different tasks and rapidly adapt to new tasks with few training samples. And meta-reinforcement learning is simply the combination of meta-learning and reinforcement learning.

Model-Agnostic Meta-Learning (MAML) is a general meta-learning algorithm proposed in \cite{finn2017model}. MAML algorithm learns an efficient initialization of the deep neural network (DNN) for different tasks through gradient descent. If there exists a distribution $p(T)$ over tasks, then some training tasks can be sampled from $p(T)$ for training the initialization $\phi$. $\phi$ is an efficient initialization if it can quickly adapt to an unseen task $T_j \sim p(T)$ with one or a few gradient descents and few training data. In this work, we apply MAML algorithm to find such an efficient initialization for similar channel selection tasks. Details of the MAML algorithm and the proposed framework will be discussed in Section IV.
\subsection{Related Work}
The method of MAML was proposed in \cite{finn2017model} and it was shown to demonstrate strong performance in both regression and deep reinforcement learning. However, training of MAML can become unstable when there is even a tiny change in the neural network structure. With this observation, the authors in \cite{antoniou2018train} proposed MAML++ algorithm which contains schemes for stabilizing the training. Another challenge is that training MAML involves second derivatives when conducting backpropagation, which increases the computational cost. In \cite{finn2017model}, the authors noted that the second derivatives can be omitted, reducing MAML to first-order MAML (FOMAML). To address the problem of high computational cost, several other first-order approximations of MAML are proposed, including Reptile in \cite{nichol2018reptile}, Hessian-free MAML (HF-MAML) in \cite{fallah2020convergence} and Evolution-Strategies MAML (ES-MAML) in \cite{song2019maml}. In our study, FOMAML is considered in the proposed framework for reducing the computational cost.

\subsection{Contributions}
Although the method of MAML keeps evolving and is being shown to exhibit strong performance in classification datasets (such as Omniglot and ImageNet) \cite{chen2019closer} and reinforcement learning applications (e.g., ant robot and 2D-navigation) \cite{finn2017model} \cite{rothfuss2018promp}, it has not been extensively applied to wireless scenarios yet. In this work, we address the wireless channel access problem in a dynamic environment. A meta-DRL framework combining MAML and deep reinforcement learning algorithms is proposed for learning a good initialization for all the similar tasks that follow the same distribution. We show that the proposed algorithm can learn a good initialization and only a few gradient descents and training data are required for adapting to the task-specific model which has near-optimal performance in the corresponding task.

We compare the proposed framework with joint-learning, which aims to find a global model for all similar tasks. The results show that there does not exist such a model that has strong performance in all the tasks and it is also difficult for the joint-learning model to adapt quickly to each specific task within a few updates.

\section{Problem Statement}
\subsection{Environment}
In this work, we consider dynamic channel access in wireless communications. There are $N$ channels for transmission. Channel conditions are represented by the vector $C=[c_1,c_2,...,c_N]$, where $c_n$ is a binary indicator of the condition of channel $n$. Channel $n$ can be in either good ($c_n=1$) or poor ($c_n=0$) condition and the transmissions over channels with poor conditions (due to deep fading or high levels of interference) fail. A time-slotted system is considered in this work, where the channel condition will be static during each time slot.

\begin{figure}[h]
	\centering
	\includegraphics[width=0.5\textwidth]{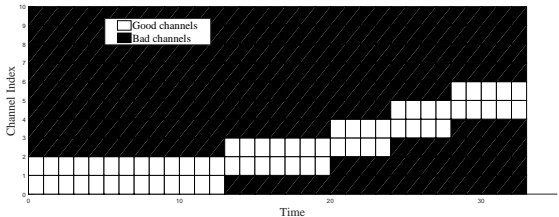}
	\caption{Channel State with $N=10$ and $p=0.1$.}
	\label{CS0.1}
\end{figure}

\begin{figure}[h]
	\centering
	\includegraphics[width=0.5\textwidth]{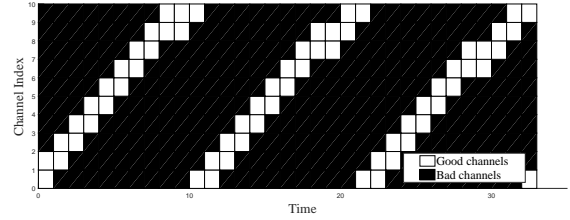}
	\caption{Channel State with $N=10$ and $p=0.9$.}
	\label{CS0.9}
\end{figure}

Similar to \cite{wang2018deep} and \cite{zhong2019deep}, it is further assumed that the channel conditions $C$ vary according to a statistical pattern. For instance, let us consider a round-robin transition pattern with two channels being in good state at a given time. In this scheme, at time $t$, assume that channels $n$ and $n+1$ are in good condition (i.e. $c_n=c_{n+1}=1$ in $C$ and the other entries remain zero). Then at time $t+1$, the next good channels will be channels $n+1$ and $n+2$ (i.e. $c_{n+1}=c_{n+2}=1$ in $C$) with probability $p$ or they will still be channels $n$ and $n+1$ with probability $1-p$. Here, $p$ is the transition probability of the round-robin pattern. Examples of round-robin patterns with two good channels among $N=10$ channels are depicted in Fig. \ref{CS0.1} and Fig. \ref{CS0.9}, with transition probabilities 0.1 and 0.9, respectively. White squares indicate channels with good conditions. In this example, we consider a simplified scenario with only two out of $N$ channels being in good condition and transitions occurring according to a statistical round-robin pattern. In a more general setting, we can have an arbitrary subset of channels being in good condition, and transitions can follow other statistical patterns, leading to another subset of channels being good conditions in the next time slot.

We consider channel access tasks with the same total number of available channels $N$ but with different transition probabilities $p$ as similar tasks. And we define that the tasks are drawn from the same distribution if their transition probabilities are drawn from the same distribution.

\subsection{Action and Reward}
In each time slot $t$, the user selects a channel (e.g., channel $n$) to transmit. If the channel is in good condition, the reward $r_t$ of taking action $a_t=n$ is 1, otherwise if the channel selected is in poor condition, the reward is $r_t=-1$.
\subsection{Observation}
Observation $O_t=[o_1,o_2,...,o_N]$ is a record of the transmission experience in time slot $t-1$. After taking action $a_t=n$ based on $O_t$ and receiving reward $r_t$, we reset all the elements in list $O_t$ to 0 and then set $o_n=r_t$ to get $O_{t+1}$. The observation can be seen as a record of the transmission in the previous time slot and it is used for decision making in the current time slot.
\subsection{Optimal Strategy}
If the statistical pattern is fixed and known, the optimal channel access strategy is as follows. If the transition probability satisfies $p>0.5$, the user will select channel $n+1$ at time $t+1$ if a transmission over channel $n$ succeeds at time $t$. On the other hand, if $p<0.5$, the user will stay in channel $n$ at time $t+1$ if a transmission over channel $n$ succeeds at time $t$. This optimal strategy implies that for different transition probabilities, there does not exist a single model that can output different strategies with the same observation as input.

Due to the randomness in channel variations, the user may also fail to transmit over channel $n$ at time $t$ with a probability of $1-p$, even if the optimal strategies above are taken. In this case, the user needs to select channel $n$ at time $t+1$ if the transition probability $p>0.5$ and select channel $n+1$ at time $t+1$ if the transition probability $p<0.5$. This strategy quickly detects the next good channel when transmission fails. The optimal strategy is summarized in Algorithm \ref{algorithm1}. Further details on the optimal strategy can be found in \cite{wang2018deep}.

\begin{algorithm}[h]
	\caption{Optimal Strategy}
	\begin{spacing}{0.8}
		\begin{algorithmic}[1]
		\STATE{At time slot $t=0$, select the good channel (action $a_0$).}
		\FOR{t=1,2,... do}
		\IF{$p<0.5$}
		\IF{channel $a_{t-1}$ is good}
		\STATE{Select channel $a_t = a_{t-1}$}
		\ELSIF{channel $a_{t-1}$ is bad}
		\STATE{Select channel $a_t = a_{t-1}+1$}
		\ENDIF
		\ELSIF{$p>0.5$}
		\IF{channel $a_{t-1}$ is good}
		\STATE{Select channel $a_t = a_{t-1}+1$}
		\ELSIF{channel $a_{t-1}$ is bad}
		\STATE{Select channel $a_t = a_{t-1}$}
		\ENDIF
		\ENDIF
		\ENDFOR
		\end{algorithmic}
	\end{spacing}
	\label{algorithm1}
\end{algorithm}

\section{Deep Reinforcement Learning}
The user/DRL agent (that performs dynamic channel access) does not know the channel pattern and it interacts with the environment by taking action $a_t$ based on its current state $s_t$. The agent gets a reward for taking the action $a_t$ and learns a policy by updating DNN parameters with the experience tuple ($s_t, a_t, r_t$).
\subsection{Agent}
Agent in this work is the user in the wireless network. The user employs DRL to determine which channel to use for successful transmission. The goal of the agent is to maximize the discounted expected sum reward in the future.
\subsection{State}
We assume the state $s_t$ is the observation $O_t$ at time $t$, in which the transmission experience in the previous time slot is recorded.
\subsection{Action and Reward}
Agent can choose one out of $N$ channels available in the wireless network. Hence, we have $a_t \in [1,2,...,N]$. Reward $r_t$ depends on the action $a_t$ and the channel conditions. If the agent chooses a channel in good state, the transmission succeeds and $r_t=1$. Otherwise, the transmission fails and $r_t=-1$.

\subsection{Policy Gradient}
We consider policy gradient method for learning the DNN parameters $\theta$. In the channel selection problem, next state $s_{t+1}$ follows the distribution $p(s_{t+1}|s_t, a_t)$ so the problem can be formulated as a Markov decision process (MDP) for a specific transition probability $p$.

With the policy gradient method, experiments are done in episodes and each episode contains $H$ time slots. During time slot $t$, action $a_t$ will be taken and then reward $r_t$ and state $s_t$ will be obtained. Each episode can be seen as an MDP with trajectory ($s_1, a_1,r_1,...s_H,a_H,r_H$). DNN parameters $\theta$ are then updated by minimizing the loss function
\begin{equation}
L(\theta)= -\frac{1}{H}\sum_{t=1}^{H}\left[ \left(\sum_{i=t}^H{\gamma^i r_i}\right)\log(\pi_{\theta}(a_t | s_t))\right],
\label{loss}
\end{equation}
where $\gamma$ denotes the discount factor in the policy gradient algorithm. $\pi_{\theta}(s_t)$ denotes the DRL policy with parameters $\theta$, which maps the current state $s_t$ to the probabilities of actions in the policy gradient algorithm. In time slot $t$, agent will take the action with the highest probability to maximize the total reward of an episode.

In (\ref{loss}), $\sum_{i=t}^H{\gamma^i r_i}$ is the expected future reward of taking action $a_t$ at state $s_t$ and $\log(\pi_{\theta}(a_t | s_t))$ is the logarithm of the probability of taking action $a_t$ at state $s_t$. After each episode, DRL agent will update once by minimizing (\ref{loss}) and learn a strategy to maximize the expected reward in the future.

\section{Meta-Reinforcement Learning}
Meta-reinforcement learning applies meta-learning to deep reinforcement learning (DRL). Since DRL is considered in this work, we refer to the proposed framework as meta-DRL. The goal of meta-DRL is to ``learn to learn". In other words, it aims to rapidly learn to perform new tasks with the assistance of previous tasks drawn from the same distribution $p(T)$. In MAML, it is required that such tasks have the same general structure, e.g., in terms of state sizes and action sizes.

MAML, proposed in \cite{finn2017model}, performs meta-learning via gradient descent. It is also compatible with any structure of DNN. MAML aims to provide a good DNN initialization $\phi$. Through only a small number of gradient descent and limited data, parameter $\theta_i$ desired in task $T_i$ can be obtained from $\phi$. In the following, we describe the approach to obtaining such an initialization $\phi$.

\begin{figure}[h]
	\centering
	\includegraphics[width=0.4\textwidth]{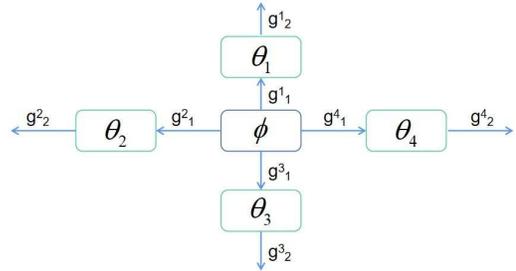}
	\caption{Diagram of MAML.}
	\label{MAML}
\end{figure}

Key operations in the MAML algorithm are depicted in Fig. \ref{MAML}. In conventional DRL, the goal is to train a task-specific model $\theta_i$, leading to strong performance in performing task $i$. However, training such a model usually requires large quantities of data collected from the environment, which may be costly or not readily available. Inspired by this, MAML aims to get an initialization $\phi$, such that through a single gradient descent $g^i$ with several MDP trajectories sampled from task $i$, the task-specific model $\theta_i$ can be reached. i.e.
\begin{equation}
\theta_i=\phi-\alpha g_1^i=\phi-\alpha\triangledown_{\phi} L_{T_i}(\phi)
\label{adapt}
\end{equation}
where $\alpha$ denotes the adaptation learning rate and $\triangledown_{\phi} L_{T_i}(\phi)$ denotes the gradient of loss over $\phi$ with the MDP trajectories sampled from task $i$. In the rest of this section, we assume that only one gradient descent is needed.

In MAML, the loss function is in the form of
\begin{equation}
L(\phi)=\sum_{i=1}^{K}L_{T_i}(\theta_i)
\label{MAMLloss}
\end{equation}
where $K$ is the number of tasks sampled from $p(T)$ in every meta-DRL iteration. $K$ is also referred as meta-batch size.

The loss function in (\ref{MAMLloss}) indicates that MAML is designed to minimize the sum loss of the evolved parameters $\theta_i$ on task $T_i$. Specifically, MAML attempts to find an initialization $\phi$ such that after one gradient descent, the $T_i$-specific models $\theta_i$ can be obtained and each $\theta_i$ can achieve the optimal performance on $T_i$.

With the goal to minimize $L(\phi)$, the update of $\phi$ can be expressed as
\begin{equation}
\phi=\phi-\beta \triangledown_{\phi}L(\phi)=\phi-\beta \triangledown_{\phi}\sum_{i=1}^{K}L_{T_i}(\theta_i)
\label{updatephi}
\end{equation}
where $\beta$ denotes the meta learning rate.

Finding the gradient in (\ref{updatephi}) is not trivial. By using the chain rule, we can rewrite the gradient as
\begin{align}
\begin{split}
\triangledown_{\phi}\sum_{i=1}^{K}L_{T_i}(\theta_i)&=\sum_{i=1}^{K}\triangledown_{\phi}L_{T_i}(\theta_i)
\\&=\sum_{i=1}^{K}\frac{\partial L_{T_i}(\theta_i)}{\partial \phi}=\sum_{i=1}^{K} \sum_{j} \frac{\partial L_{T_i}(\theta_i)}{\partial \theta_i^j}\frac{\partial \theta_i^j}{\partial \phi}
\end{split}
\label{gradient}
\end{align}
where $\theta_i^j$ denotes the $j^{th}$ parameter in $\theta_i$.

Now, (\ref{adapt}) can be rewritten as
\begin{equation}
\theta_i^j=\phi^j-\alpha\frac{\partial L_{T_i}(\phi)}{\partial \phi^j}.
\label{adapt2}
\end{equation}

Therefore, each element of the last term $\frac{\partial \theta_i^j}{\partial \phi}$ in (\ref{gradient}) can be computed as
\begin{equation}
\frac{\partial \theta_i^j}{\partial \phi^m}=
\left\{\begin{matrix}
-\alpha \frac{\partial L_{T_i}(\phi)}{\partial \phi^j \partial \phi^m}& j\neq m\\
1-\alpha \frac{\partial^2 L_{T_i}(\phi)}{\partial {\phi^j}^2}& j=m.
\end{matrix}\right.
\label{hessian}
\end{equation}

In \cite{finn2017model}, the authors proposed the first-order MAML (FOMAML), where the Hessian matrices in (\ref{hessian}) are ignored, i.e. the value of (\ref{hessian}) becomes 0 if $j \neq m$ and 1 if $j=m$. The work in \cite{nichol2018reptile} provides the proof and numerical results showing that FOMAML can achieve similar performance levels as MAML and reduce the computation cost dramatically. In our work, we employ FOMAML in the proposed meta-DRL framework for dynamic channel access.

With the approximation in FOMAML, gradient in (\ref{gradient}) is simplified as
\begin{equation}
\sum_{i=1}^{K} \triangledown_{\phi} L_{T_i}(\theta_i) =\sum_{i=1}^{K} \frac{\partial L_{T_i}(\theta_i)}{\partial \theta_i}= \sum_{i=1}^{K} \triangledown_{\theta_i} L_{T_i}(\theta_i).
\label{simpleGradient}
\end{equation}

In the next section, we discuss in detail how to implement (\ref{simpleGradient}) and determine a good initialization $\phi$ in meta-DRL.

\section{Meta-DRL Framework for Dynamic Channel Access}
\subsection{Meta-DRL Algorithm}
In prior work in \cite{wang2018deep} and \cite{zhong2019deep}, a single model is trained for a specific task $T_i$ with the fixed transition probability $p_i$. Such a model can then achieve strong performance in the same task $T_i$. However, training such a task-specific model is expensive and the model will fail when $p_i$ varies (due to e.g., changes in the statistics of the channel conditions, interference, other users' channel access strategies, resource allocation mechanisms, etc.). In this work, we propose the meta-DRL framework in order to find a good initialization $\phi$ for the DNN parameters for different transition probabilities $p_i$, starting from which the DRL agent can quickly adapt to the optimal channel selection strategies for an arbitrary $p_i$.

After applying the approximation in FOMAML, the gradient in (\ref{simpleGradient}) is simply the sum or average of the second-update gradients in each task $T_i$, i.e. $g_2^i$ as shown in Fig. \ref{MAML}. Details of the implementation are provided in Algorithm \ref{algorithm2} below. Although we present the derivation with only two inner updates, we can have several more inner updates according to \cite{finn2017model}. For instance, if $I$ inner updates are performed, the first $I-1$ updates aim to obtain the task-specific parameter $\theta_i$ and the last update is the evaluation of $\theta_i$ in task $T_i$. FOMAML will collect the last gradients from all $T_i$ for updating the initialization $\phi$.

After obtaining the initialization $\phi$ via Algorithm \ref{algorithm2}, we randomly select a new task from the same distribution and perform a few gradient descents in the new task with the neural network initialized as $\phi$. Then, the performance of the post-updated parameters in the new task will be used for evaluating $\phi$.

\begin{algorithm}[h]
	\caption{Meta-DRL}
	\begin{spacing}{0.8}
		\begin{algorithmic}[1]
			\STATE{Initialize $\phi=\phi_0$}
			\WHILE{not done}
			\STATE{Sample a batch of channel selection tasks $T_i \sim p(T)$ with probabilities $p_i$ respectively.}
			\FORALL{$T_i$}
			\STATE{Sample an episode of experiences $D=(s_1,a_1,r_1,...,s_H,a_H,r_H$) in task $T_i$ using DNN parameters $\phi$.}
			\STATE{Perform gradient descent using experience $D$ and compute the $T_i$-specific parameters $\theta_i$, as shown in (\ref{adapt}).}
			\STATE{Sample a new episode of experiences $D'=(s_1,a_1,r_1,...,s_H,a_H,r_H$) in task $T_i$ using DNN parameters $\theta_i$.}
			\STATE{Perform gradient descent using experience $D'$ and obtain $\theta_i^{'}$.}
			\ENDFOR
			\STATE{Update $\phi \leftarrow \phi- \frac{\beta}{K} \sum_{i=1}^{K} \triangledown_{\theta_i} L_{T_i}(\theta_i)=\phi- \frac{\beta}{K} \sum_{T_i}(\theta_i^{'}-\theta_i)$}.
			\ENDWHILE
		\end{algorithmic}
	\end{spacing}
	\label{algorithm2}
\end{algorithm}

\begin{algorithm}[h]
	\caption{Joint-Learning}
	\begin{spacing}{0.8}
		\begin{algorithmic}[1]
			\STATE{Initialize $\phi=\phi_0$}
			\WHILE{not done}
			\STATE{Sample a batch of channel selection tasks $T_i \sim p(T)$ with probabilities $p_i$ respectively.}
			\FORALL{$T_i$}
			\STATE{Sample an episode of experiences $D=(s_1,a_1,r_1,...,s_H,a_H,r_H$) in task $T_i$ using DNN parameters $\phi$.}
			\STATE{Perform gradient descent using experience $D$ and update $\phi\leftarrow \phi-\alpha_2 \triangledown_{\phi} L_{T_i}(\phi)$.}
			\ENDFOR
			\ENDWHILE
		\end{algorithmic}
	\end{spacing}
	\label{algorithm3}
\end{algorithm}

\begin{algorithm}[h]
	\caption{Sampling experiences in each episode}
	\begin{spacing}{0.8}
		\begin{algorithmic}[1]
			\STATE{Assuming the agent has already occupied a good channel at $t=0$ and has a corresponding obeservation $O_1$, hence it starts at a good state $s_1=O_1$.}
			
			\FOR{$t=1$ to $H$}
			\STATE{Channel state evolves following round-robin.}
			\STATE{The agent will select an action $a_t$ with the DNN and state $s_t$.}
			\STATE{After transmission, the agent will get access to the observation $O_{t+1}$ which reflects the condition of the selected channel.}
			\STATE{$s_{t+1}=O_{t+1}$.}
			\ENDFOR
			
		\end{algorithmic}
	\end{spacing}
	\label{algorithm4}
\end{algorithm}

\subsection{Benchmark Algorithms}
In this work, we compare meta-DRL with joint-learning. Joint-learning aims at finding a global model that performs well on all the tasks. Implementation of joint-learning is explained in Algorithm \ref{algorithm3}. For a fair comparison, joint-learning uses a similar algorithmic structure as FOMAML. The only difference is that there is only a single update performed in each sampled task $T_i$ and this gradient will be directly applied in updating the neural network $\phi$.

\section{Numerical Results and Analysis}
\subsection{Experimental Setup}
\begin{table}[h]
	\renewcommand{\arraystretch}{1.3}
	
	\caption{Experimental Parameters}
	\label{table1}
	\centering
	\small
	\begin{tabular}{|c||c|}
		\hline Number of Channels $(N)$ & 10\\
		\hline Number of Good Channels  & 2\\
		\hline Meta-batch Size ($K$) & 15\\
		\hline Episode Length ($H$) & 30\\
		\hline Number of Inner Updates & 15\\
		\hline Adaptation Learning Rate ($\alpha$) & 0.1\\	
		\hline Meta Learning Rate ($\beta$)	& 0.05\\
		\hline Discounted Factor ($\gamma$) & 0.9\\
		\hline
		
	\end{tabular}
	\label{table}
\end{table}

Hyperparameters used in the proposed meta-reinforcement learning framework are listed in Table \ref{table}. For the task distribution, the transition probability are selected from $p \in [0,0.2] \cup [0.8,1]$. We consider these transition probability values because the channel conditions will change almost randomly with $p$ around 0.5, which is not practical and will not contribute to the training.

Initially, 100 tasks with different transition probabilities are randomly sampled from the distribution. Then, for every meta-DRL iteration, we sample a meta-batch with 15 tasks from the 100 available tasks and perform FOMAML following Algorithm \ref{algorithm2}. For each inner update and adaptation update, we average the gradients obtained from 20 sampled episodes. It is necessary to evaluate each gradient with multiple episodes to ensure the reliability of each update.

Each episode starts with a random good channel, which prevents the channels from constantly staying in the same conditions with small transition probabilities. It is also assumed that the agent has selected one of the two good channels at the beginning of each episode and hence it starts with a good initial state.

For both meta-DRL and joint learning, we use a DNN with two layers consisting of 50 neurons and 20 neurons, respectively. ReLU is used as the activation function and Adam optimizer is employed for minimizing the loss function during the inner updates.

Meta-batch size and the two learning rates $\alpha$, $\beta$ are some of the critical hyperparameters. $\alpha$ needs to be as large as possible so that DNN can quickly adapt to different tasks. On the other hand, too large a value for $\alpha$ will also lead to performance collapse during the adaption. Meta-batch size should be large enough to make the initialization $\phi$ to generalize over the tasks. However, large meta-batch size will also dramatically increase the training time.

\subsection{Numerical Results}

Every 100 iterations during the training of meta-DRL, we evaluate the performance of $\phi$ by utilizing it in 50 validation tasks with different transition probabilities randomly selected from the set $[0,0.2] \cup [0.8,1]$. Utilizing $\phi$ in a task means that $\phi$ is used as the initialization of DNN and 20 gradient descents are performed in the task for adaptation.

In Fig. 3, we conduct 2000 training iterations and compare the performances of meta-DRL and joint-learning. Meta-DRL attains the performance of the average transmission success rate (SR) after 20 updates on the validation tasks. Note that transmission is successful if a channel with good state is selected. Hence, transmission success rate is the same as the rate of accessing channels in good state, which vary according to a statistical pattern. At the start of the training, it can be seen that 20 updates are insufficient with random DNN initialization and can only achieve an average SR of approximately 50.8\%. After meta-DRL converges at around 2000 iterations, it can achieve an average SR of around 81.1\%. For the considered distribution of $p$, transmissions can fail even if the agent learns the optimal strategy. This is due to the randomness of the round-robin channel pattern and the ideal average SR for $p \sim [0,0.2] \cup [0.8,1]$ is 90\%. Hence we conclude that meta-DRL achieves a near-optimal performance in all the tasks with $p \sim [0,0.2] \cup [0.8,1]$.

\begin{figure}[h]
	\centering
	\includegraphics[width=0.4\textwidth]{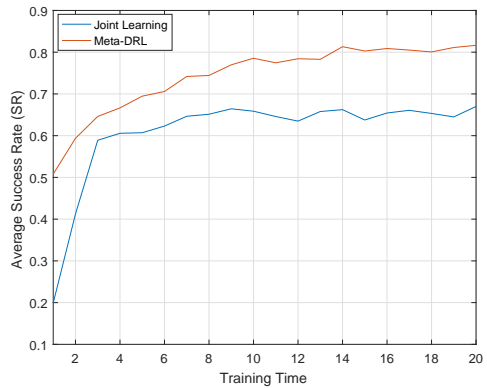}
	\caption{Validation on 50 Sampled Tasks during Meta-DRL Training.}
	\label{Training}
\end{figure}

The best performance of joint-learning is achieved with learning rate $\alpha_2=0.001$, while the rest of the hyperparameters remain the same as in meta-DRL. In Fig. \ref{Training}, joint-learning only achieves an SR of around 66\% after convergence. Its performance implies that there does not exist a global DNN that can handle all the tasks with different transition probabilities. At the end of Section II, we have noted that, for the same observation, different strategies should be pursued under different transition probabilities, and this cannot be accomplished with a single DNN. And FOMAML based meta-DRL learns an initialization for quickly adapting to different task-specific DNNs and hence outperforms joint-learning.

\begin{figure}[h]
	\centering
	\includegraphics[width=0.4\textwidth]{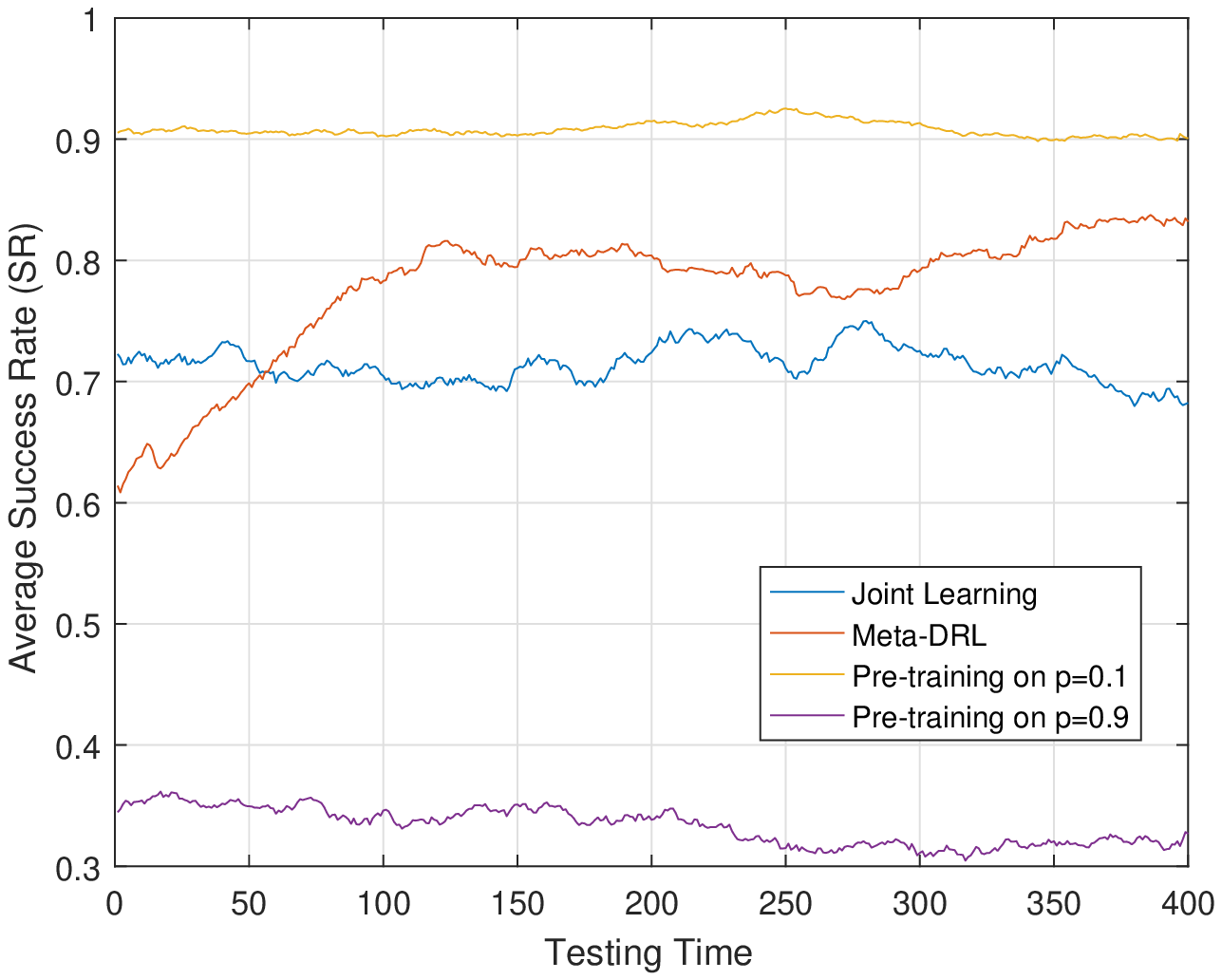}
	\caption{Adaptation in Channel Selection Task with p=0.1.}
	\label{0.1test}
\end{figure}

\begin{figure}[h]
	\centering
	\includegraphics[width=0.4\textwidth]{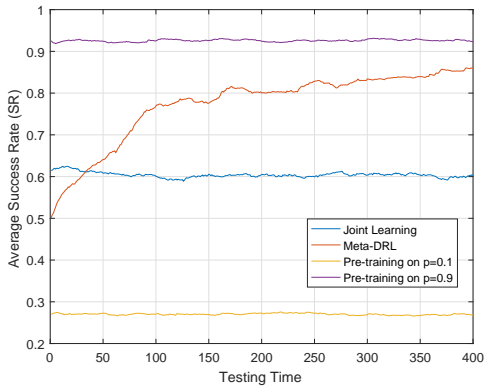}
	\caption{Adaptation in Channel Selection Task with p=0.9.}
	\label{0.9test}
\end{figure}

We further examine the efficiency of the initialization $\phi$ for a given individual task. We denote the parameters obtained in joint-learning as $\phi'$. In this part, two tasks with the transition probabilities $p$ of 0.1 and 0.9, respectively, are examined. We assign $\phi$ and $\phi'$ to the neural networks and let them perform 20 updates in each of the tasks. Fig. \ref{0.1test} and Fig. \ref{0.9test} depict the performance during 20 updates (400 episodes) in the tasks. We observe in these figures that although $\phi$ itself has a poor performance in different tasks, it can quickly adapt to task-specific parameters $\theta_i$ leading to a strong performance in task $i$. On the other hand, $\phi'$ obtained in joint-learning has a strong performance at the beginning because joint-learning aims at finding a global model that performs as strong as possible over all tasks. However, it cannot adapt to a specific task within a small number of updates.

Furthermore, we compared meta-DRL with the pre-training in the previous work in \cite{wang2018deep} and \cite{zhong2019deep}, where a task-specific model is trained for a fixed transition probability. The DNN consists of two layers with 50 neurons each and the learning rate is set to be 0.001. Two converged DNNs are obtained for $p=0.1$ and $p=0.9$ respectively. We then let them update in both tasks and the performance during adaptation is shown in Fig. \ref{0.1test} and Fig. \ref{0.9test}. It can be seen that the task-specific DNNs achieve the optimal performance in the corresponding task, however, they fail to adapt to the task with a different probability.

We select the best performance out of different adaptation learning rates for the benchmarks and the numerical results imply that none of them can adapt to different tasks drawn from the same distribution. It can be concluded that the proposed meta-DRL framework not only learns the transition pattern of the channels but also estimates the transition probabilities within a few updates and limited training data.

\section{Conclusions and Future Work}
DRL is now being intensively studied in wireless communications due to its beneficial features of being model-free and learning without prior knowledge. However, training neural networks require a huge amount of data, which can become costly or impractical for wireless systems due to the sheer size of many different wireless states and environments. Therefore, learning similar tasks with the method of meta-DRL will dramatically decrease the demand for training data and computational resources.

In this work, we have proposed a meta-DRL framework to address the dynamic channel access problem in wireless communications with time-varying channel conditions. Instead of learning a task-specific model that can only perform well in the corresponding task, we have designed a meta-DRL agent that learns an efficient initialization for similar tasks. Numerical results show that the joint-learning algorithm fails to adapt to similar tasks. Besides, we show that although the task-specific model can achieve optimal performance in the corresponding task, it fails to adapt to a new task. On the other hand, with the efficient initialization learned by meta-DRL, near-optimal performance for dynamic channel access in different tasks drawn from the same distribution can be achieved with only a few updates and limited training data.


\bibliographystyle{IEEEtran}
\bibliography{ref}
\end{document}